\documentclass[floatfix,superscriptaddress,twocolumn,notitlepage]{revtex4-1}
\pdfoutput=1
\usepackage{graphicx,latexsym}
\usepackage{amssymb,amsthm,amsmath}
\usepackage{cancel}
\usepackage{epstopdf}
\usepackage[utf8]{inputenc}
\usepackage{placeins}

\begin{document}

\title{Quantitative Chemically-Specific Coherent Diffractive Imaging of Buried Interfaces using a Tabletop EUV Nanoscope}
\author{Elisabeth R. Shanblatt}
\email{elisabeth.shanblatt@colorado.edu}
\affiliation{These authors contributed equally to this work.}
\affiliation{JILA, University of Colorado, 440 UCB, Boulder, Colorado 80309-0440, USA}
\author{Christina L. Porter}
\email{christina.porter@colorado.edu}
\affiliation{These authors contributed equally to this work.}
\affiliation{JILA, University of Colorado, 440 UCB, Boulder, Colorado 80309-0440, USA}
\author{Dennis F. Gardner}
\affiliation{JILA, University of Colorado, 440 UCB, Boulder, Colorado 80309-0440, USA}
\author{Giulia F. Mancini}
\affiliation{JILA, University of Colorado, 440 UCB, Boulder, Colorado 80309-0440, USA}
\author{Robert M. Karl Jr.}
\affiliation{JILA, University of Colorado, 440 UCB, Boulder, Colorado 80309-0440, USA}
\author{Michael D. Tanksalvala}
\affiliation{JILA, University of Colorado, 440 UCB, Boulder, Colorado 80309-0440, USA}
\author{Charles S. Bevis}
\affiliation{JILA, University of Colorado, 440 UCB, Boulder, Colorado 80309-0440, USA}
\author{Victor H. Vartanian}
\affiliation{SUNY Poly SEMATECH, 257 Fuller Road, Suite 2200, Albany, NY 12203, USA}
\author{Henry C. Kapteyn}
\affiliation{JILA, University of Colorado, 440 UCB, Boulder, Colorado 80309-0440, USA}
\author{Daniel E. Adams}
\affiliation{JILA, University of Colorado, 440 UCB, Boulder, Colorado 80309-0440, USA}
\author{Margaret M. Murnane}
\affiliation{JILA, University of Colorado, 440 UCB, Boulder, Colorado 80309-0440, USA}

\date{March 2016}

\begin{abstract}

Characterizing buried layers and interfaces is critical for a host of applications in nanoscience and nano-manufacturing. Here we demonstrate non-invasive, non-destructive imaging of buried interfaces using a tabletop, extreme ultraviolet (EUV), coherent diffractive imaging (CDI) nanoscope. Copper nanostructures inlaid in SiO$_2$ are coated with 100 nm of aluminum, which is opaque to visible light and thick enough that neither optical microscopy nor atomic force microscopy can image the buried interfaces. Short wavelength (29 nm) high harmonic light can penetrate the aluminum layer, yielding high-contrast images of the buried structures. Moreover, differences in the absolute reflectivity of the interfaces before and after coating reveal the formation of interstitial diffusion and oxidation layers at the Al--Cu and Al--SiO$_2$ boundaries. Finally, we show that EUV CDI provides a unique capability for quantitative, chemically-specific imaging of buried structures, and the material evolution that occurs at these buried interfaces, compared with all other approaches. 

\end{abstract}

\maketitle

\section{Introduction}

Probing and characterizing nanoscale interfaces buried beneath visibly opaque materials is a critical capability for nanoscience and nanotechnology. Most imaging modalities cannot be used for non-destructive, sub-surface ($>$50 nm) imaging. Optical microscopy and atomic force microscopy (AFM) image the surfaces of visibly opaque samples. Backscattered electron (BSE) scanning electron microscopy (SEM), and secondary electron SEM with secondary electrons generated from backscattered electrons (SE-II)~\cite{Bernstein2013}, can image buried features with increasing electron energy providing increased penetration depth~\cite{Niedrig1998}. However, there is a tradeoff in terms of decreased resolution~\cite{Erlandsen1999}, charging of insulating samples, sample damage, and hydrocarbon buildup~\cite{Goldstein2003}. Furthermore, SEM often produces a complicated mixture of morphology and composition information making quantitative image analysis difficult. Finally, in BSE-SEM, elemental contrast for neighboring elements is often subtle because backscattered electron efficiency is proportional to the natural logarithm of the atomic number~\cite{Lloyd1987}. 

Various specialized imaging techniques have been employed for buried layer metrology. Mode-synthesizing AFM and scanning near-field ultrasound holography can detect subsurface structures~\cite{Vitry2015, Shekhawat2005}. Three-dimensional structures have also been probed using large-scale synchrotron X-ray diffraction microscopy in a transmission geometry~\cite{Song2008, Jiang2013}. However, these approaches are not easily extendable to high-throughput imaging, especially for thick objects. 

Short wavelength extreme ultraviolet (EUV) and X-ray light have unique potential for imaging buried interfaces because they can penetrate through optically opaque materials, provide chemically-specific contrast, and also image nanoscale features. In particular, by combining coherent, short-wavelength beams from either high harmonic generation (HHG)~\cite{Bartels2002} or X-ray free electron lasers (XFELs)~\cite{Chapman2009} with coherent diffractive imaging (CDI)~\cite{Sayre1952, Fienup1982, Miao1999, Robinson1999}, it is now possible to reach near-wavelength resolution imaging in the EUV and X-ray regions for the first time~\cite{Seaberg2014, Zhang2015}. Accordingly, CDI has found a range of applications in transmission and reflection geometries~\cite{Harada2013, Roy2011, Sun2012} to investigate nanoscale strain~\cite{Harder2013, Robinson2009}, semiconductor structures~\cite{Harada2013}, and for biological imaging~\cite{Jiang2010, Seibert2011, Dumas2013}. EUV/X-ray CDI can be non-destructive and suffers no charging effects or resolution loss with depth. Moreover, the contrast mechanisms in EUV CDI are relatively straightforward and intrinsically high, with amplitude images showing material composition, and phase images showing material composition and topography.

Here, we perform tabletop, high harmonic spectro-nanoscopy of buried interfaces opaque to visible light, enabling unique, non-destructive investigation of many interfacial phenomena. We use ptychographic CDI~\cite{Thibault2008, Maiden2009}, which utilizes redundant information from multiple diffraction patterns recorded with overlapping fields of view to robustly reconstruct both the amplitude and phase of a buried layer. Ptychographic CDI~\cite{Maiden2009} has been implemented in a high-NA reflection geometry using HHG illumination beams~\cite{Seaberg2014, Zhang2015}, with the amplitude of the reconstructed images yielding the relative reflectivity (or transmissivity) between different regions of the sample. Here, we also demonstrate a modification of ptychography that returns absolute reflectivities simply by measuring the flux of the illuminating beam. This new capability called Reconstructed Absolute Ptychographic Transmissivity/Reflectivity CDI (RAPTR-CDI) allows us to non-destructively detect the formation of interstitial diffusion and oxidation layers at the Al--Cu and Al--SiO$_2$ boundaries. We verify the presence of interdiffusion using destructive Auger electron spectroscopy sputter depth profiling (AES). Thus, we demonstrate that  EUV CDI provides a unique capability for quantitative, chemically-specific imaging of buried structures. This will make possible unprecedented studies of material evolution at buried interfaces, compared with all other approaches. In the future, the fast temporal resolution of high harmonics can be harnessed for imaging dynamically functioning and evolving interfaces.

\section{Experiment}

\begin{figure}[h!]
\begin{center}
\includegraphics[width= 8.6cm]{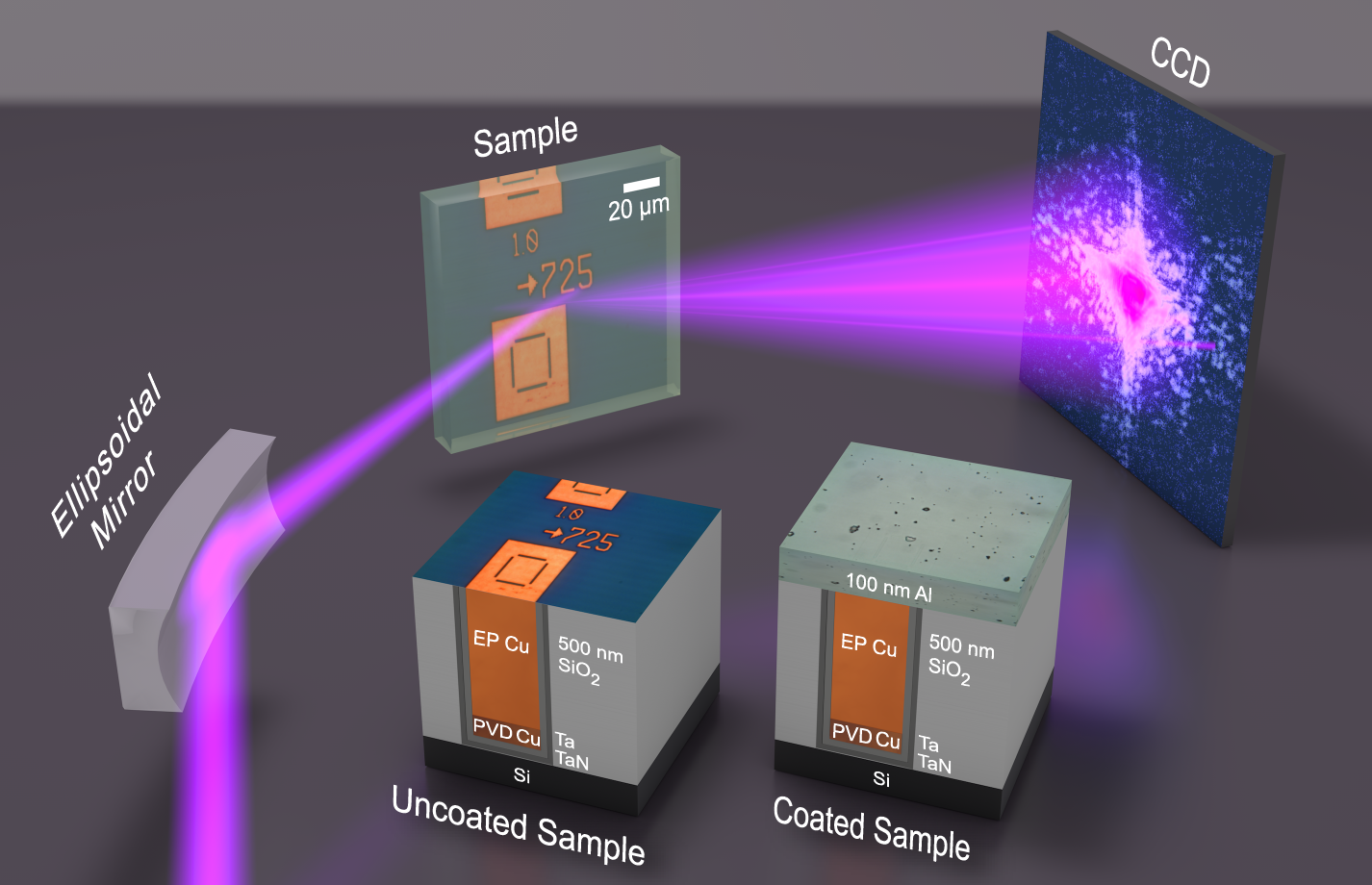}
\caption{\textbf{Schematic of the Experiment and Samples.} Schematic of the experimental setup. The inset shows a diagram of the two samples imaged, topped with their optical microscope images. The tantalum was deposited by physical vapor deposition (PVD), and the top layer of copper was electroplated (EP).\label{Fig1}}
\end{center}
\end{figure}

The experimental setup is illustrated in Fig.\,\ref{Fig1}. Bright, phase-matched, fully spatially coherent high harmonic beams~\cite{Corkum1993, Rundquist1998, Bartels2002, Popmintchev2010} were generated by focusing a Ti:Sapphire laser beam (23 fs, 1.5 mJ, 785 nm pulses at 5 kHz) into a 5 cm-long waveguide filled with Argon gas at 49 Torr~\cite{Bartels2002, Popmintchev2010}. Harmonics around the $27^\text{th}$ order (29.1 nm) were reflected from two super-polished silicon substrates set at Brewster’s angle to reflect the HHG beam while rejecting residual laser light. Two \hbox{200 nm-thick} Al filters were used to block any remaining fundamental laser light. The HHG beam was then passed through an iris to induce a hard edge on the beam incident on the sample. Two narrow-bandwidth multilayer mirrors set at 12$^\circ$ and $47.7^\circ$, and with a bandpass of approximately \hbox{1 nm}, were used to select the \hbox{29.1 nm} harmonic light. An ellipsoidal mirror at a $5^\circ$ angle of incidence from the surface focused the HHG beam to a $\approx16$ $\mu$m diameter spot, which was incident on the sample at 57.8$^\circ$ from the normal. A $2048\times2048$ Princeton Instruments (PI-MTE) CCD, placed a distance of \hbox{3.85 cm} downstream of the sample and normal to the un-diffracted beam, was used to collect the light diffracted from the sample. 

The two samples imaged in this work were cleaved from a damascene-style wafer consisting of Cu structures inlaid in SiO$_2$~\cite{Hummler2013}, provided by SEMATECH. The wafer was polished flat using chemical mechanical planarization (CMP). After exposure to atmospheric conditions for 14 months, repeating patterned cells were cleaved such that essentially identical areas of interest would be present on two samples. One of the samples was coated with 100 nm of Al using an Edwards Cryo 304 electron-gun physical vapor deposition system, while the other was left uncoated for comparison. For both samples, the ptychographic data set consisted of 270 diffraction patterns, collected with 3 $\mu$m step sizes between scan positions. A random offset of $\pm$20\% of the step size was added to each scan position to prevent periodic artifacts in the reconstructions~\cite{Thibault2009}. The total EUV exposure time for the uncoated sample was 5.8 min, compared with 23.6 min for the Al-coated sample, with a total scanned area of 4270 $\mu$m$^2$ each. In future experiments, these exposure times can be reduced significantly ($>$10$\times$) using optimized driving lasers.

\begin{figure*}[htb!]
\begin{center}
\includegraphics[width= 6in]{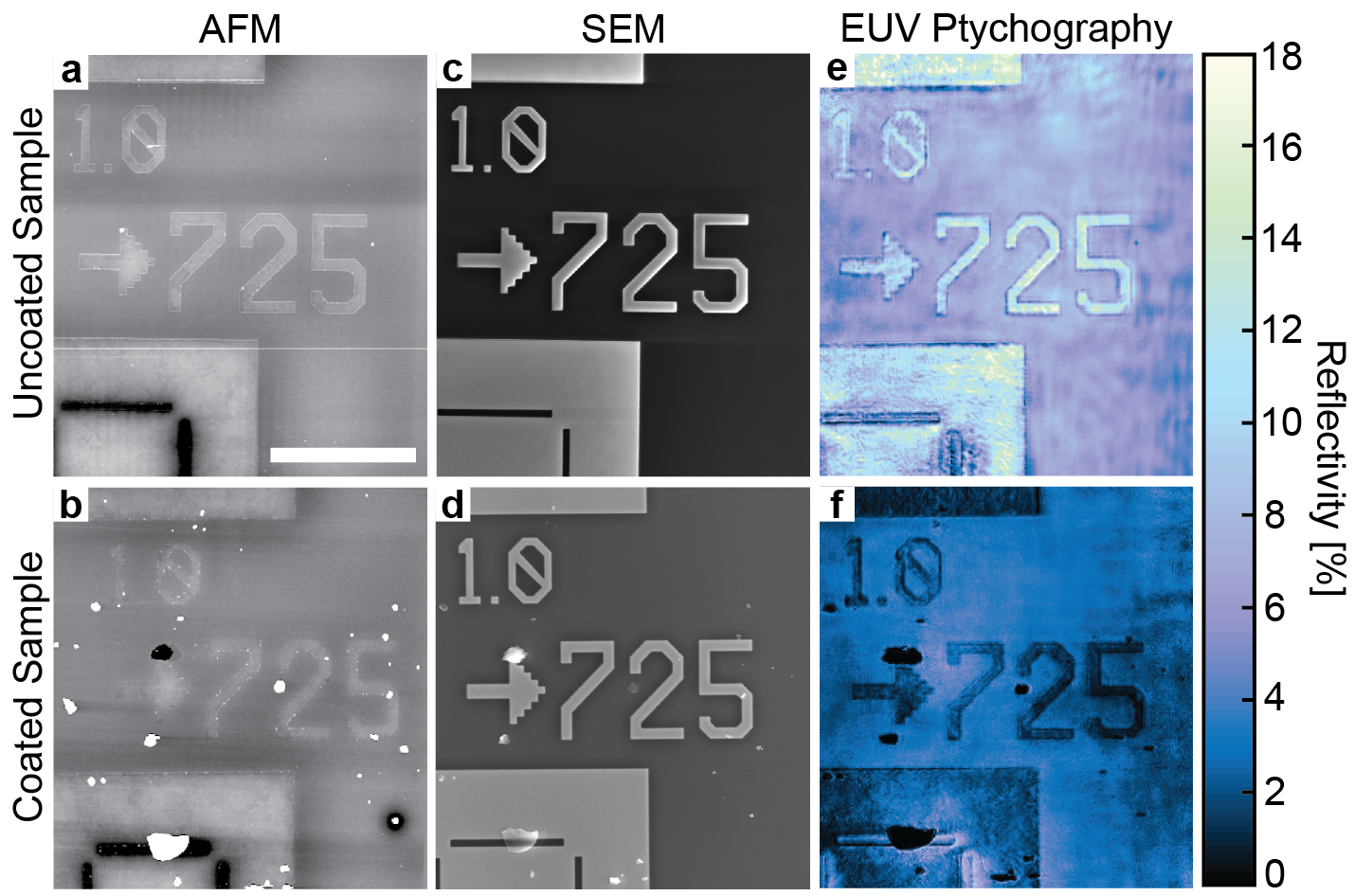} 
\caption{\textbf{Comparison of AFM, SEM, and EUV Ptychography Images.}
Uncoated sample (top row) and Al-coated sample (bottom row). AFM images with a 20 $\mu$m scale bar are shown in (a,b) with black corresponding to \hbox{0 nm} height and white to \hbox{60 nm} height. The AFM images were flattened with 5$^\text{th}$ order polynomial fits to remove substrate curvature and horizontal lines were removed where the AFM tip erroneously jumped due to surface contamination. SEM images collected with a secondary electron detector at \hbox{18 kV} are shown in (c,d). EUV ptychography images obtained with the multicolor refinement algorithm are shown in (e,f). The reconstructions are scaled to show absolute reflectivity based on the mean absolute reflectivity of the features and substrate in the RAPTR-CDI reconstructions. \label{Fig2}}
\end{center}
\end{figure*}

For both samples, the object and probe were computationally reconstructed using a combination of ptychography algorithms. The diffraction patterns were interpolated onto a linear spatial frequency grid using tilted plane correction~\cite{Gardner2012}, then reconstructed with RAPTR-CDI with position correction~\cite{Zhang2013}. The multilayer mirrors allowed small amounts of adjacent harmonics to leak through, broadening the total fractional bandwidth of the illumination and degrading the fidelity of the reconstructions. For this reason, the corrected positions were next fed into a multicolor ptychographic information multiplexing algorithm~\cite{Batey2014}. By using this multicolor algorithm, noise due to the presence of the unwanted harmonics could be filtered out so that only 29.1 nm light contributed to the reconstructed images. The multicolor algorithm significantly improved the image fidelity over single color ptychography (see Supplementary Fig.~S1). 

In order to use RAPTR-CDI to characterize the absolute reflectivity of the buried metal-metal and metal-oxide interfaces at 29.1 nm, we measured the flux of the HHG beam before each ptychography scan. We recorded images of the beam reflected from a gold mirror, then used the exposure time and reflectivity of gold to calculate the number of available photons incident on the sample, in units of detector counts. The reflectivity of the gold mirror was taken to be 27.9\%~\cite{Windt1998} with the surface roughness assumed to be that of sibling silicon substrates used in Ref.~\cite{Zhang2015}. Then, in RAPTR-CDI, the probe is normalized to the measured power in each iteration. The absolute value squared of the reconstructed object’s complex amplitude is therefore equal to the sample’s reflected intensity at every pixel. Because the sample reflectivity is spatially varying, masks for the features and substrates (Supplementary Fig.~S2) were used to select regions free from contamination. We segmented the RAPTR-CDI reconstructions to calculate the average absolute reflectivities reported in Fig.~\ref{Fig4}. We also scaled the multicolor reconstructions such that the mean reflectivity of the masked regions agrees with that of the RAPTR-CDI images. The scaled multicolor reconstructions for both samples are shown in Fig.~\ref{Fig2}. 

\section{Chemically Specific Imaging of Buried Metal-Metal and Metal-Oxide Interfaces}

\begin{figure*}[htb!]
\begin{center}
\includegraphics[width= 7in]{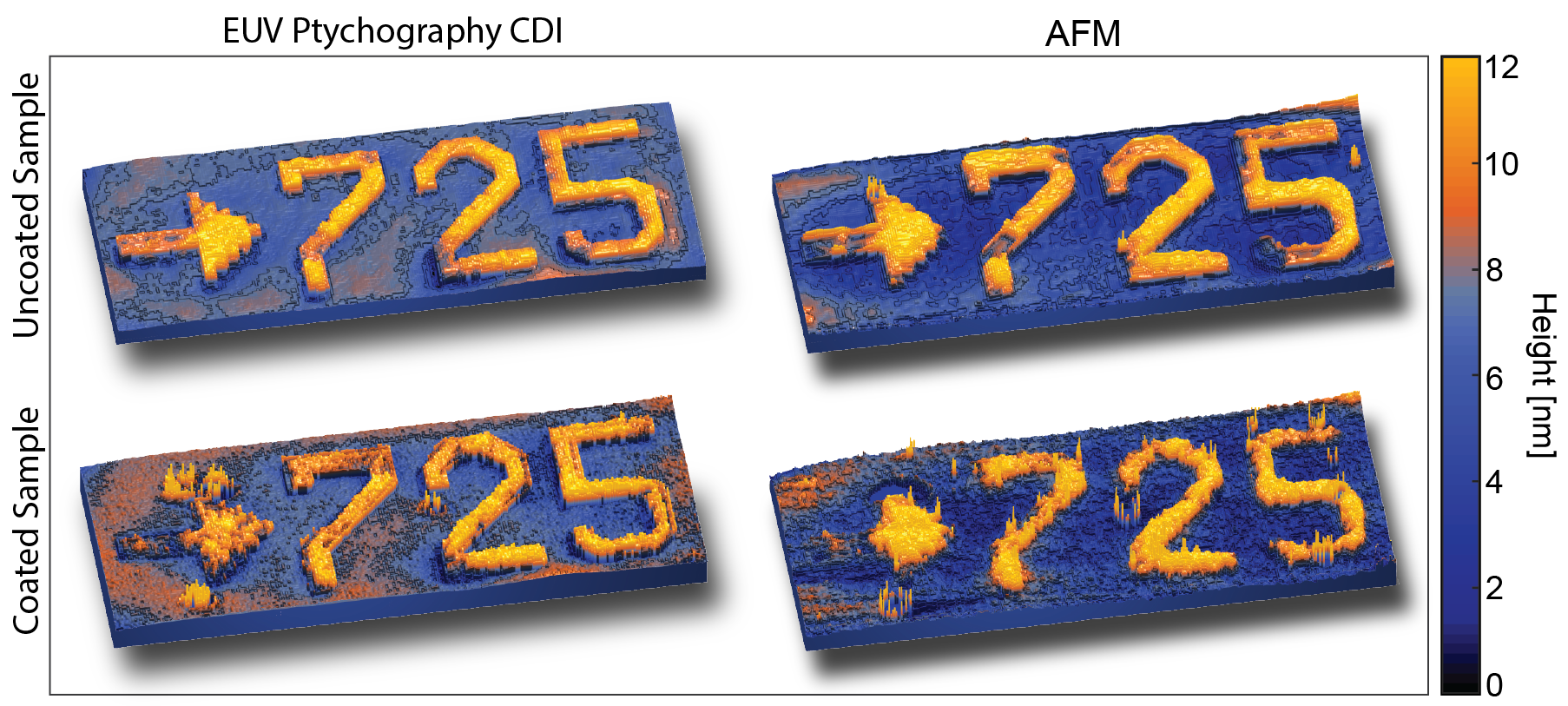} 
\caption{\textbf{Comparison of EUV CDI Height Maps with AFM.} Height maps for the uncoated and Al-coated samples. Both the coated EUV and AFM images show high aspect-ratio artifacts where debris is located on the surface of the sample.\label{Fig3}}
\end{center}
\end{figure*}

To highlight the extreme sensitivity of EUV light to interfacial composition, we compare our multicolor CDI amplitude images (which have been normalized to the measured RAPTR-CDI reflectivities) to optical microscopy (Fig.~\ref{Fig1}) as well as secondary electron SEM and AFM (Fig.~\ref{Fig2}). The SEM (FEI Nova NanoSEM 630) with an Everhart-Thornley secondary electron detector and an accelerating voltage of 18 kV shows contrast on both samples (Fig.~\ref{Fig2}c,d). The contrast on the coated sample is due to a combination of morphology on the surface of the Al as well as the chemical difference between the underlying features and substrate. This elemental contrast is detectable because the Everhart-Thornley detector also captures SE-II electrons generated from backscattered electrons reflected from the underlying structures~\cite{Goldstein2003}. This means that the coated SEM image includes a complex mixture of height and material information that is not easily decoupled.

The AFM (Digital Instruments MMAFM-2) only shows surface topography. The coated features are visible on the top surface because Al deposition is slow (5 Å, or approximately two atomic layers, per second), causing each layer to acquire the underlying topography. However, the AFM’s accuracy in the vicinity of surface contamination is significantly degraded due to the much higher aspect ratio of the contamination than the nearly-flat features. 

The EUV ptychography amplitude reconstruction (Fig.~\ref{Fig2}f) is the only image that definitively visualizes the buried interfaces, as confirmed by the amplitude contrast between the Cu features and substrate. Indeed, if the EUV nanoscope were only imaging the top Al surface, then the absolute reflectivity image in Fig.~\ref{Fig2} would appear featureless and the reconstruction would be phase-only. Instead, EUV light penetrates through the Al to reveal the buried structures, yielding the observed reflectivity variation. Additionally, the penetration depth of Al at 29.1 nm exceeds 400 nm, meaning that 80\% of the light is transmitted through 100 nm of Al.
The transverse spatial resolution of the EUV images is limited by the effective numerical aperture (NA) of the imaging system. Because there are no optics between the sample and detector, the NA is determined by the distance from the sample to the detector ($z = 38.5$ mm) and the size of the detector (D). The diffraction patterns were cropped to $512\times512$ to reduce computation time, so the effective detector size, $D_\text{eff}$ is $512 \times p$, with pixel size \hbox{$p = 13.5\,\mu$m} square. Therefore, the NA is $D_\text{eff}/z = 0.09$ and the diffraction limited resolution is \hbox{$\lambda/(2NA) = 162$ nm.}

\section{High Resolution Surface Topography from Phase Reconstructions}

In addition to the chemical composition discernible from the amplitude of the reconstructions (shown in Fig.~\ref{Fig2}), the phase of the reconstructions contains both material and height information. By subtracting the phase of the complex reflectivity predicted by our modeled feature and substrate stacks (discussed below) from the reconstructed phase images, we can generate height maps showing the surface topography of the samples~\cite{Seaberg2014, Zhang2015}. In the case of the coated sample, the phase of the exit surface wave accounts for phase changes within the stack, and the height map therefore shows the top surface topography as opposed to the buried surface. We compare these height maps to AFM height maps shown in Fig.~\ref{Fig3}. The z-axis is scaled by a factor of 200 to highlight the height variations between the features and substrate. There is very good agreement between the height maps generated from ptychogrpahy and AFM.

\section{Quantitative Detection of Reactions and Diffusion at Buried Interfaces}

In order to compare our experimental reflectivity measurements to those theoretically predicted, we calculated the complex reflectivities of multilayer stacks representative of the features and substrate for the damascene samples, using methods derived from Ref~\cite{Moharam1995}, which solve Maxwell's equations directly in the multilayer stack. We confirmed our predictions with IMD~\cite{Windt1998}, a software program that uses the Fresnel equations to calculate the complex reflectivity for a stack of materials at EUV/X-ray wavelengths~\cite{Nevot1980, Stearns1989}. Our multilayer stack method and IMD agree to within 0.3\% reflectivity and $5^\circ$ reflected phase, which would correspond to a 0.38 nm height difference in a height map. 

\begin{figure}[htb!]
\begin{center}
\includegraphics[width=8.6cm]{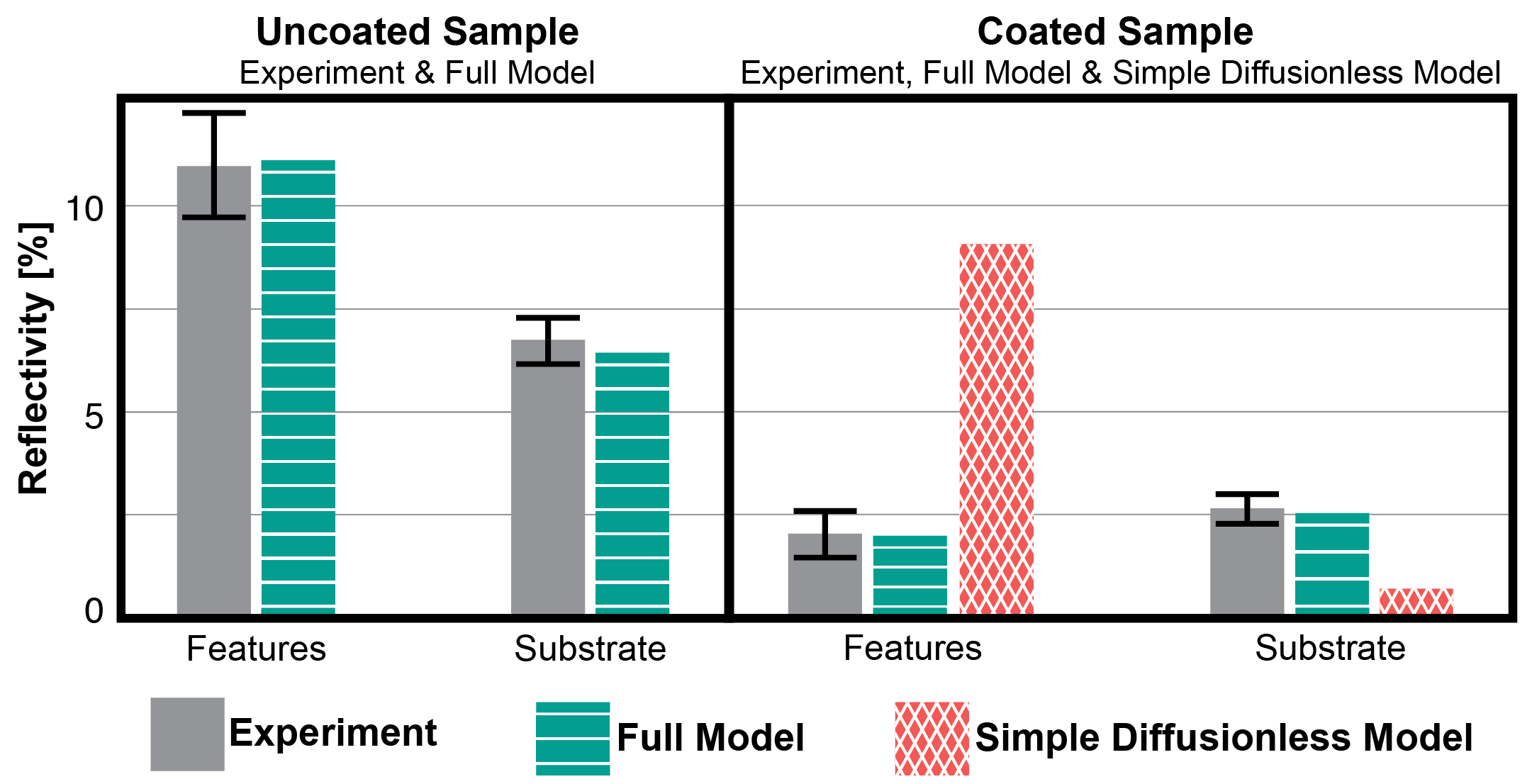} 
\caption{\textbf{Experimental vs.~Modeled Reflectivities.} Experimentally measured reflectivities compared to modeled reflectivities for the uncoated and Al-coated samples. For the coated sample, the green bars represent predictions by a model that includes diffusion, and the red bars represent predictions by a model with the same material quantities as those in the full model, but with no diffusion. Because the primary sources of uncertainty in these calculations arise from the oxide layer thicknesses, which are intrinsically difficult to quantify, we omit error bars from our model results. \label{Fig4}}
\end{center}
\end{figure}

The four distinct areas—uncoated substrate, uncoated features, coated substrate, and coated features—are modeled as multilayer stacks with values for surface roughness, oxide layer thickness, composition, and interdiffusivity informed by AFM and AES, as discussed below. Modeling the propagation of light through these stacks returns a complex reflectivity coefficient. We verify that the intensity matches our measured reflectivity, while the phase produces height maps consistent with the AFM. Therefore, the modeled stacks have three constraints: the AES data (for applicable regions), the reconstructed intensity, and the reconstructed phase. The uncertainty in the experimental measurement includes uncertainty in the gold mirror reflectivity and the standard deviation from the mean reflectivity due to spatial variation in the reconstructed reflectivity. Our final reflectivity models (green striped bars) are compared to the experimentally measured reflectivities (grey bars) in Fig.~\ref{Fig4}.

In the case of the uncoated sample, we estimate that the Cu features form a native oxide bilayer upon exposure to the atmosphere consisting of 4.8 nm CuO atop 7.4 nm Cu$_2$O, based on a study of copper oxide growth at ambient temperatures indicating a ratio of $0.65 = d_\text{CuO}/d_{\text{Cu}_2\text{O}}$~\cite{Keil2007}. We calculate the total oxide thickness by assuming that the sample was polished flat during chemical mechanical planarization and the average feature height measured by the AFM is all due to oxide growth. This total thickness agrees with the amount of oxide predicted based on the AES measurement of the coated features (discussed below). We assume the substrate is simply SiO$_2$. These modeled stacks lead to good agreement for the reflectivity of the uncoated sample.

To predict the coated sample feature reflectivity, we initially used a simple model with 100 nm of Al deposited on the same stacks used for the uncoated sample. This model led us to predict $R_\text{features} = 5.6$\% and $R_\text{substrate} = 3.8$\%.  However, these predictions are significantly higher than both of our experimental measurements (which were $R_\text{features} = 2.0$\% and $R_\text{substrate} = 2.7$\%). Furthermore, the reconstruction displays a contrast inversion in the coated sample image, where the coated Cu features are less reflective than the coated substrate, whereas they were more reflective in the uncoated sample (Fig.~\ref{Fig2}e,f) and in the simple model for the coated sample. Adding an Al$_2$O$_3$ layer on top of the Al layer could not resolve this contrast inversion, nor could assuming that the Al scavenged O from the copper oxides to form Al$_2$O$_3$ at the interface, which is an energetically favorable reaction since 2Al + 3CuO $\rightarrow$ Al$_2$O$_3$ + 3Cu corresponds to $\Delta_f$H$^\circ = -1207.52$ kJ/mol~\cite{Chase1998}. 

\begin{figure}[htb!]
\begin{center}
\includegraphics[width=7cm]{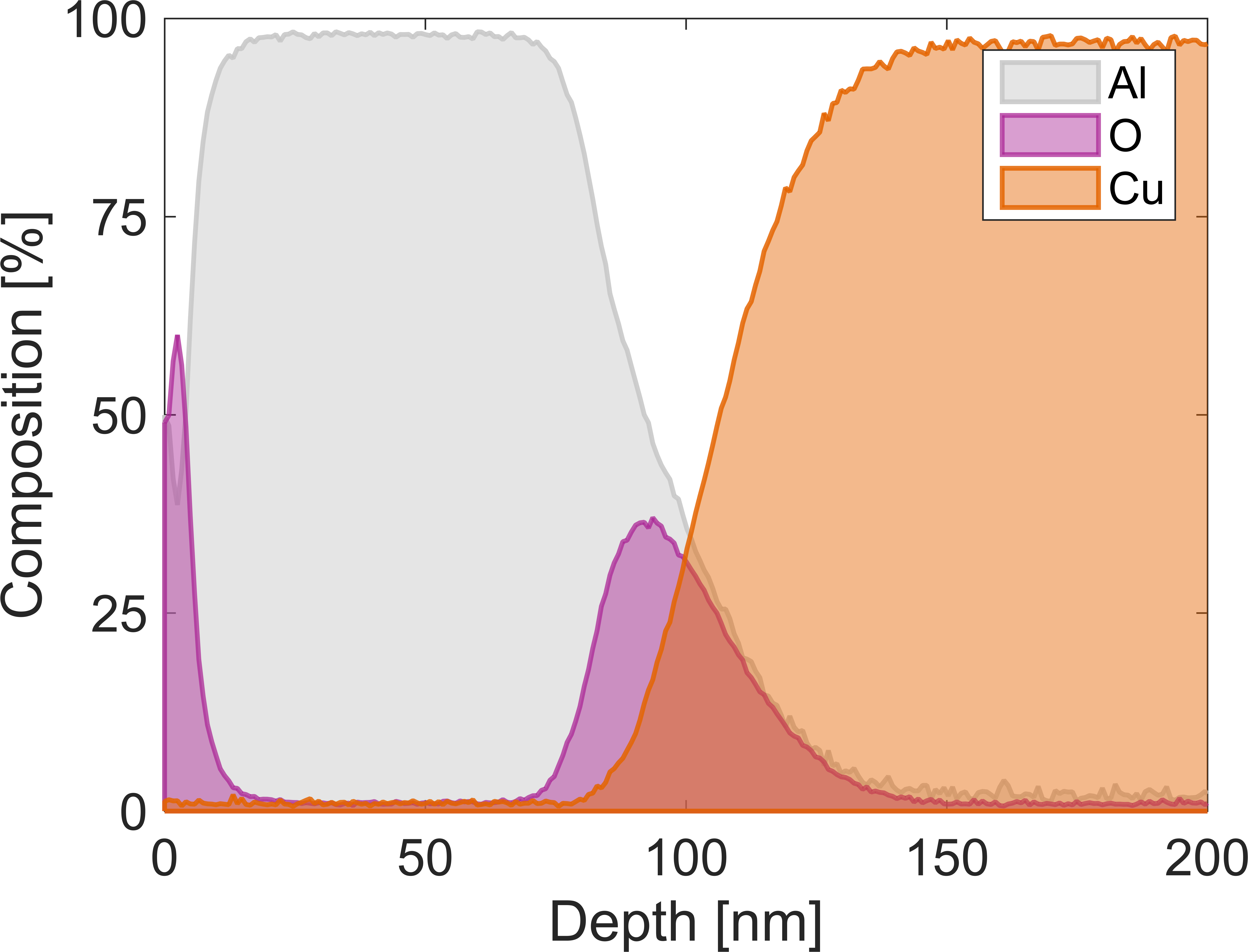} 
\caption{\textbf{Auger electron sputter depth profile.} Auger electron spectroscopy sputter depth profile of a coated sample feature, showing a diffusive region between the Al and Cu. \label{Fig5}}
\end{center}
\end{figure}

The seeming inexplicability of this relative contrast inversion led us to hypothesize that Kirkendall diffusion occurred at the Al--Cu boundary, forming an interstitial layer that decreased the reflectivity at the interface~\cite{Nakajima1997}. We confirmed the presence of this interstitial diffusion layer on the coated features using AES sputter depth profiling, a destructive technique that requires ion sputtering through the sample to obtain Auger electron spectra at every relevant depth~\cite{Hofmann2014} (Fig.~\ref{Fig5}). This technique revealed a 40 nm diffusive region at the interface. We used the AES depth profile shown in Fig.~\ref{Fig5} to calculate the theoretical reflectivity of the Al-coated features and substrate. In the feature regions, we calculated stoichiometrically plausible percent compositions for Al, Cu, Al$_2$O$_3$, CuO, and Cu$_2$O from the elemental composition provided by the AES at each depth. We tested two diffusive models, one in which Al$_2$O$_3$ is the only oxide at the Al--Cu interface and one in which there is a mixture of Al$_2$O$_3$, CuO, and Cu$_2$O at the interface. We find that the model in which we assume Al captures all of the O from the copper oxides at the boundary fits well with our measured reflectivity (Fig.~\ref{Fig4}).  

\begin{figure*}[ht!]
\begin{center}
\includegraphics[width=6.5in]{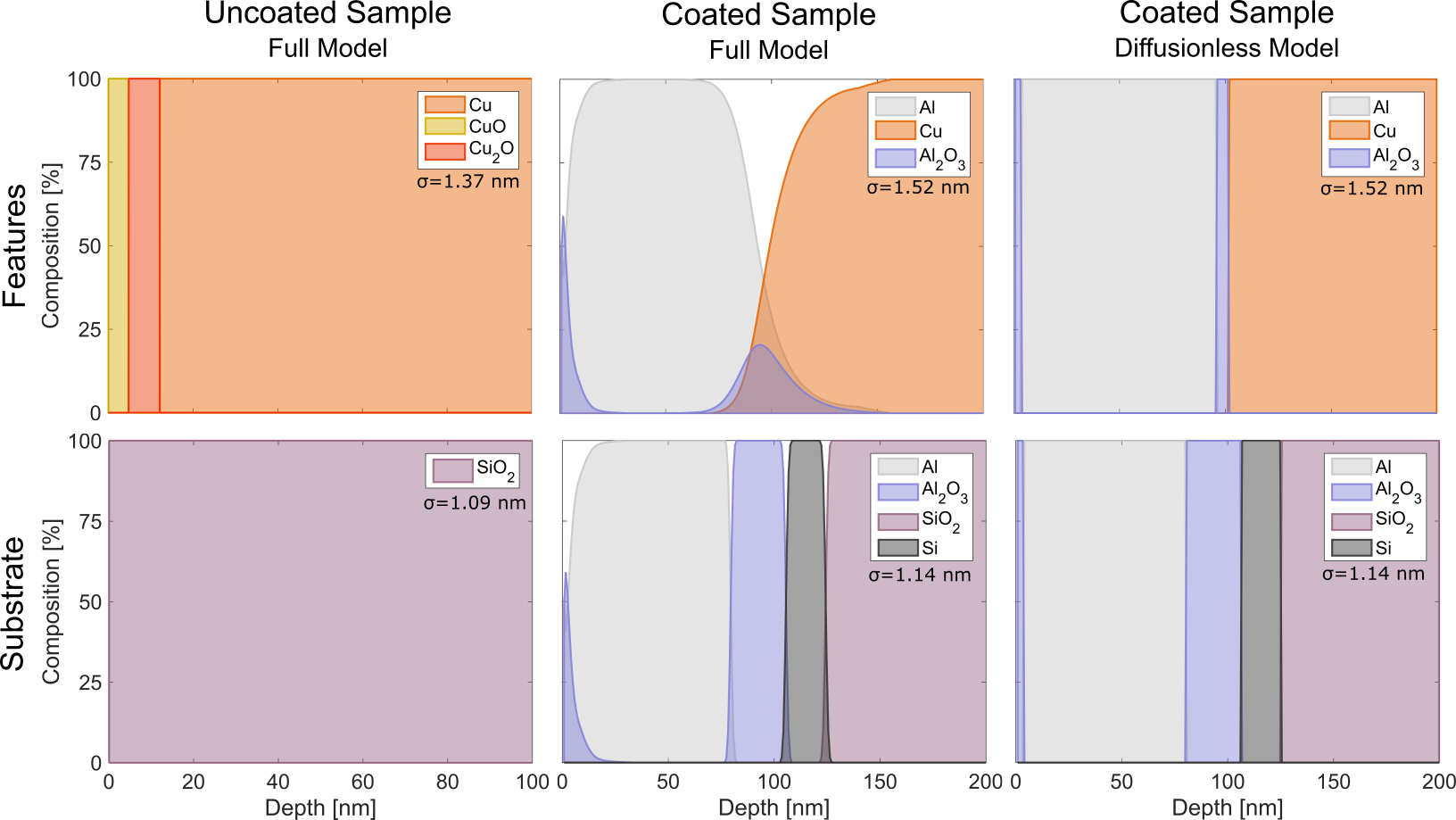}  
\caption{\textbf{Depth profiles used to model sample reflectivities.} Stacks used to model the reflectivity of different parts of the samples and their surface roughnesses. The top row shows the stacks corresponding to the features, and the bottom row shows stacks corresponding to the substrate. The full model for the features includes significant diffusion. Neglecting this diffusion (shown in the third column) predicts reflectivity values highly inconsistent with the reconstructed reflectivities. \label{Fig6}}
\end{center}
\end{figure*}

For the coated substrate, we used the same stack as in the full model of the coated features for the top 50 nm of the sample (including a 3 nm diffusive Al$_2$O$_3$ layer), followed by an abrupt Al--SiO$_2$ interface at a depth corresponding to 100 nm of deposited Al. However, this resulted in a predicted reflectivity of 7.4\%, which is twice as bright as we observe experimentally. We resolved this discrepancy by incorporating 26.5 nm of Al$_2$O$_3$ at the Al--SiO$_2$ interface, with diffusion modeled by convolving the depth profile with a Gaussian of full-width 6 nm. Since this Al$_2$O$_3$ layer corresponds to Al scavenging O from SiO$_2$, we incorporated the corresponding amount of liberated silicon in a layer below the Al$_2$O$_3$. This modification produced agreement between the experimental reflectivity (2.66\%) and the model (2.57\%). We were unable to experimentally confirm the presence of the diffusive Al$_2$O$_3$ layer at the interface because SiO$_2$ is non-conductive, preventing a reliable AES depth profile on the coated substrate. However, there is experimental evidence in the literature indicating that Al can scavenge oxygen from SiO$_2$ when deposited via e-beam evaporation~\cite{Roberts1981, Zeng1983}. Visual representations of all the stacks used in our models, including surface roughnesses, are shown in Fig.\,\ref{Fig6}. 

To investigate how sensitive reflectivity in the EUV spectral range is to diffusion, we modeled our final coated sample stacks (Fig.~\ref{Fig6}, middle column) with the same amount of material as in our full model, but without any diffusion (Fig.~\ref{Fig6}, right column). The predictions of this diffusionless model are shown by the red crosshatched bars in the right panel of Fig.~\ref{Fig4}. They dramatically disagree with our measurements, indicating that our method is highly sensitive to diffusion. 

\section{Conclusions}

We combined ptychographic CDI with EUV high harmonic beams to demonstrate a unique new capability for quantitatively imaging buried interfaces with chemically-specific contrast through metal layers that are opaque to visible light, AFM, and SE-I SEM. We developed a new technique, RAPTR-CDI, which is a modified ptychography algorithm that yields absolute reflectivities by normalizing the probe at every iteration to the correct measured flux incident on the sample. This allowed us to non-destructively detect the existence of interstitial diffusion and oxidation layers between the buried Cu and SiO$_2$ structures and the Al coating. The sample features and interdiffusion layer thicknesses agree well with those characterized using AFM, SEM, Auger electron spectroscopy as well as values from the literature~\cite{Keil2007}. We also demonstrated the utility of multicolor ptychography as a spectral filter to improve resolution and image quality compared to single color reconstructions, even with only minor contributions from neighboring harmonics.

In the future, we can extend this work by observing interfacial charge, energy and spin transport, examining the effect of increased temperature on interstitial diffusion, and using a comb of harmonics spanning an absorption edge to simultaneously image through multiple layers. RAPTR-CDI also opens the door to fully quantitative material characterization combined with EUV imaging, much like imaging ellipsometry, by performing angle-resolved measurements with the sample placed on a tilt stage. This would allow for highly-sensitive metrology of thick samples inaccessible to optical ellipsometry. The technique could further be extended to perform high-throughput buried layer imaging by using multiple EUV beams to achieve a wide field of view~\cite{Karl2015}.

\section{Methods}

\subsection{Sample Fabrication}
The Cu structures were deposited by physical vapor deposition and patterned using I-line resist with a Canon GS22 followed by electroplating to reach a total height of \hbox{750 nm}. Subsequently, \hbox{500 nm} of SiO$_2$ was deposited using tetraethyl orthosilicate (TEOS).

\subsection{Uncoated Sample Reflectivity Model Calculation}

\begin{flushleft} It is possible that a small amount of interdiffusion may occur between the Cu and its oxide layers, but we found that modeling diffusion by convolving the depth profile with a Gaussian having up to a \hbox{5 nm} standard deviation did not significantly alter the theoretical reflectivity or phase.\\
\end{flushleft}
\begin{samepage}
\begin{table}[h]
\begin{ruledtabular}
    \centering
    \begin{tabular}{ccc}
                 & Features &  Substrate  \\ \hline
      Experiment & $10.97 \pm 1.26$      &  $6.77 \pm 0.61$        \\
      Model      & 11.10                 & 6.45
    \end{tabular}
    \caption{Experimentally reconstructed mean reflectivity values (\%) compared to modeled reflectivities of features and substrate on the uncoated sample. The error shown is the standard deviation from the mean of the reconstructed reflectivity in the masked regions (Fig.\,S2). \label{UncoatedTable}}
\end{ruledtabular}
\end{table}
\end{samepage}

\subsection{Coated Sample Reflectivity Values}
\FloatBarrier
\begin{table}
\begin{ruledtabular}
    \centering
    \begin{tabular}{ccc}
                 & Features &  Substrate  \\ \hline
      Experiment & $2.04 \pm 0.56$      &  $2.66 \pm 0.35$        \\
      Model      & 1.99                 & 2.57
  
    \end{tabular}
    \caption{Experimentally reconstructed mean reflectivity (\%) values compared to modeled reflectivities of features and substrate on the coated sample.  \label{CoatedTable}}
\end{ruledtabular}
\end{table}
\FloatBarrier

\subsection{Height Maps}
In the case of the uncoated sample, the model predicts the Fresnel phase shift due to the features is -96$^\circ$, while the phase change from the substrate is -65$^\circ$. In the case of the coated sample, the phase change from the features is -92$^\circ$, while the phase change from the substrate is -43$^\circ$.

For plotting purposes, the AFM images have been smoothed with a Gaussian filter with full width equal to the EUV microscope’s pixel size in the case of the coated sample and twice the EUV pixel size in the case of the uncoated sample (to remove a one pixel artifact in the image around the letters). Without smoothing, the AFM images appear noisy because the feature heights are similar to the surface roughness of the sample. 

Images were flattened using a $5^\text{th}$ order polynomial surface fit to decrease substrate height variations and make the feature heights more easily comparable. Binary masks used to assign Fresnel phases in the CDI reconstructions were generated using the intelligent scissors algorithm~\cite{Mortensen1995}.

\subsection{Ptychography Reconstructions}
Each of the reconstructions were run first using RAPTR-CDI for an initial 50 iterations, with position correction~\cite{Zhang2013} implemented for another 3000 iterations afterwards. The multicolor reconstructions were both run using the wavelengths of the 25$^\text{th}$, 27$^\text{th}$, and 29$^\text{th}$ harmonics and for approximately 2000 iterations. 

The difference between the positions found using the position correction algorithm and the initial recorded positions were used to correctly determine the pixel size of the reconstructed images. This is necessary in general for ptychography CDI because any error in the measurement of the sample to detector distance or angle results in changes in the reconstructed pixel size. To solve for this change in the $x$ and $y$ directions, the difference between the corrected positions and the initial positions was fit to a plane. In particular, the scale factor necessary to correct the pixel size in the $x$-direction, $a_x$ is $1-s_x$, where $s_x$ is the slope of the plane fit to the position differences. Then, $d_x\cdot a_x$ is the corrected pixel size with $d_x$ being the predicted pixel size in the absence of tilted plane correction and any error in the measured detector-sample distance. It was found that after solving for the correct pixel size the scaling of the EUV reconstructions agreed well with the SEM measurements, whereas without this scaling the reconstructions were stretched by up to 18\% in a given direction.

To correct for error in the XY-calibration of the AFM used for comparison to the HHG CDI reconstructions, the AFM images in Figs.~\ref{Fig2} and~\ref{Fig3} were scaled in the horizontal and vertical directions so that the transverse dimensions of the number ``725" are in agreement with the SEM images shown in Fig.~\ref{Fig2}.

\begin{figure*}[h!]
\begin{center}
\includegraphics[width=6in]{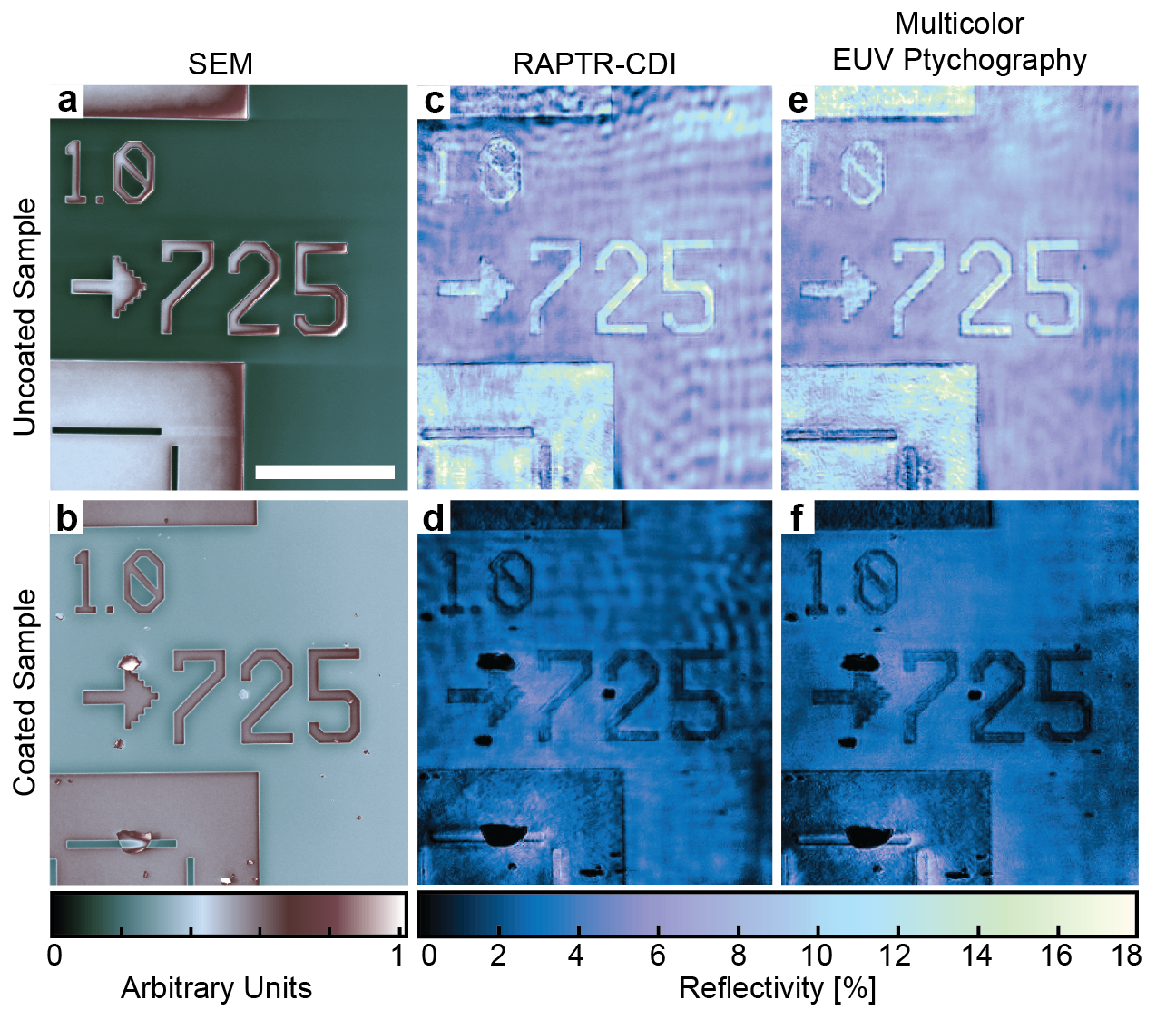} 
\caption{\textbf{RAPTR-CDI vs. multicolor reconstructions.} 
(a,b) SEM images for comparison. (c,d) RAPTR-CDI reconstructions, which return absolute reflectivity, but assume all diffraction is from exactly one harmonic. (e,f) Multicolor ptychography reconstructions demonstrating the enhanced fidelity provided by the multicolor algorithm. Because the multicolor algorithm only returns relative reflectivity, the reconstructions have been scaled such that their average reflectivity agrees with that of the RAPTR-CDI reconstructions. 
\label{FigS1}}
\end{center}
\end{figure*}

\begin{figure*}[h!]
\begin{center}
\includegraphics[width=4in]{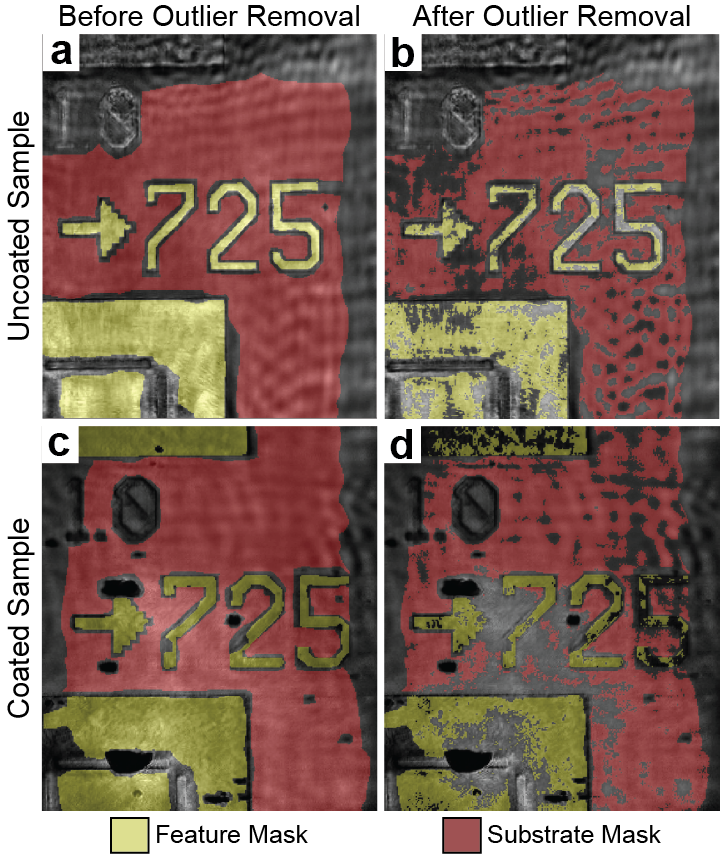} 
\caption{\textbf{Masks used in reflectivity calculations.} Masks used to calculate the average feature reflectivity (yellow) and substrate reflectivity (red) for the uncoated (b) and coated (d) sample are shown in the right column. Masks (b) and (d) were generated from masks (a) and (c) by excluding points further than one standard deviation from the mean reflectivity of the included areas. Masks (a) and (c) were made using the intelligent scissors algorithm \cite{Mortensen1995}. Masks (b) and (d) were used to scale the multicolor reconstructions so that their average reflectivity matches the RAPTR-CDI reconstructions. \label{FigS2}}
\end{center}
\end{figure*}

\clearpage

\section{Acknowledgments}
The authors would like to gratefully acknowledge support for this work from an NSF MRSEC grant DMR-1420620, the DARPA PULSE program, and a Gordon and Betty Moore Foundation EPiQS Award. We thank SEMATECH for providing the samples used in this work, David Alchenberger for helpful discussion and assistance with processing and characterization techniques, Patrick Stevens and Marilyn Andrews from Rocky Mountain Laboratories for performing the Auger electron spectroscopy and helpful discussion, as well as David Windt, Jonas Gertsch, R. Joseph Kline, and Galen Miley for helpful discussion. C. L. Porter and D. F. Gardner were supported by the NSF Graduate Research Fellowship Program and C. L. Porter was also supported by the Katharine Burr Blodgett Fellowship. E. R. Shanblatt and D. F. Gardner were supported by the NSF COSI IGERT program. D. F. Gardner was also supported by a Ford fellowship. R. M. Karl Jr. was supported by the NSF MRSEC grant DMR-1420620. G.F. Mancini was supported by the Swiss National Science Foundation through the Early Postdoc.Mobility Fellowship No.~P2ELP2\_158887.

\section{Author Contributions}
In alphabetical order: H.C.K. and M.M.M. conceived of the experiment. D.F.G. designed and constructed the experimental apparatus. D.E.A., D.F.G., R.M.K., G.F.M., C.L.P., and E.R.S. conducted the experiment. D.E.A., C.L.P., and E.R.S. performed the data analysis. V.H.V. provided the sample. All authors commented on the results and prepared the manuscript.

\section{Competing Financial Interests}
The authors declare no competing financial interests.


\clearpage

\begin{thebibliography}{10}
\expandafter\ifx\csname url\endcsname\relax
  \def\url#1{\texttt{#1}}\fi
\expandafter\ifx\csname urlprefix\endcsname\relax\def\urlprefix{URL }\fi
\providecommand{\bibinfo}[2]{#2}
\providecommand{\eprint}[2][]{\url{#2}}

\bibitem{Bernstein2013}
\bibinfo{author}{Bernstein, G.}, \bibinfo{author}{Carter, A.} \&
  \bibinfo{author}{Joy, D.}
\newblock \bibinfo{title}{Do {SE-(II)} electrons really degrade {SEM} image
  quality?}
\newblock \emph{\bibinfo{journal}{Scanning}} \textbf{\bibinfo{volume}{35},}
  \bibinfo{pages}{1--6} (\bibinfo{year}{2013}).

\bibitem{Niedrig1998}
\bibinfo{author}{Niedrig, H.} \& \bibinfo{author}{Rau, E.}
\newblock \bibinfo{title}{Information depth and spatial resolution in {BSE}
  microtomography in {SEM}}.
\newblock \emph{\bibinfo{journal}{Nucl. Instrum. Meth. B}}
  \textbf{\bibinfo{volume}{142},} \bibinfo{pages}{523--534}
  (\bibinfo{year}{1998}).

\bibitem{Erlandsen1999}
\bibinfo{author}{Erlandsen, S.}, \bibinfo{author}{Macechko, P.} \&
  \bibinfo{author}{Frethem, C.}
\newblock \bibinfo{title}{High resolution backscatter electron {(BSE)} imaging
  of immunogold with in-lens and below-the-lens field emission scanning
  electron microscopes}.
\newblock \emph{\bibinfo{journal}{Scanning Microscopy}}
  \textbf{\bibinfo{volume}{13},} \bibinfo{pages}{43--54}
  (\bibinfo{year}{1999}).

\bibitem{Goldstein2003}
\bibinfo{author}{Goldstein, J.} \emph{et~al.}
\newblock \emph{\bibinfo{title}{Scanning Electron Microscopy and {X-ray}
  Microanalysis}} (\bibinfo{publisher}{Springer Science}, \bibinfo{address}{New
  York}, \bibinfo{year}{2003}), \bibinfo{edition}{3rd} edn.

\bibitem{Lloyd1987}
\bibinfo{author}{Lloyd, G.}
\newblock \bibinfo{title}{Atomic number and crystallographic contrast images
  with the {SEM}: A review of backscattered electron techniques}.
\newblock \emph{\bibinfo{journal}{Mineralogical Mag.}}
  \textbf{\bibinfo{volume}{51},} \bibinfo{pages}{3--19} (\bibinfo{year}{1987}).

\bibitem{Vitry2015}
\bibinfo{author}{Vitry, P.} \emph{et~al.}
\newblock \bibinfo{title}{Mode-synthesizing atomic force microscopy for {3D}
  reconstruction of embedded low-density dielectric nanostructures}.
\newblock \emph{\bibinfo{journal}{Nano Research}} \textbf{\bibinfo{volume}{8},}
  \bibinfo{pages}{1--7} (\bibinfo{year}{2015}).

\bibitem{Shekhawat2005}
\bibinfo{author}{Shekhawat, G.} \& \bibinfo{author}{Dravid, V.}
\newblock \bibinfo{title}{Nanoscale imaging of buried structures via scanning
  near-field ultrasound holography}.
\newblock \emph{\bibinfo{journal}{Science}} \textbf{\bibinfo{volume}{310},}
  \bibinfo{pages}{89--92} (\bibinfo{year}{2005}).

\bibitem{Song2008}
\bibinfo{author}{Song, C.} \emph{et~al.}
\newblock \bibinfo{title}{Nanoscale imaging of buried structures with elemental
  specificity using resonant {X-ray} diffraction microscopy}.
\newblock \emph{\bibinfo{journal}{Phys. Rev. Lett.}}
  \textbf{\bibinfo{volume}{100},} \bibinfo{pages}{025504}
  (\bibinfo{year}{2008}).

\bibitem{Jiang2013}
\bibinfo{author}{Jiang, H.} \emph{et~al.}
\newblock \bibinfo{title}{Three-dimensional coherent {X-ray} diffraction
  imaging of molten iron in mantle olivine at nanoscale resolution}.
\newblock \emph{\bibinfo{journal}{Phys. Rev. Lett.}}
  \textbf{\bibinfo{volume}{110},} \bibinfo{pages}{205501}
  (\bibinfo{year}{2013}).

\bibitem{Bartels2002}
\bibinfo{author}{Bartels, R.} \emph{et~al.}
\newblock \bibinfo{title}{Generation of spatially coherent light at extreme
  ultraviolet wavelengths}.
\newblock \emph{\bibinfo{journal}{Science}} \textbf{\bibinfo{volume}{297},}
  \bibinfo{pages}{376--378} (\bibinfo{year}{2002}).

\bibitem{Chapman2009}
\bibinfo{author}{Chapman, H.}
\newblock \bibinfo{title}{{X-ray} imaging beyond the limits}.
\newblock \emph{\bibinfo{journal}{Nature Mater.}} \textbf{\bibinfo{volume}{8},}
  \bibinfo{pages}{299--301} (\bibinfo{year}{2009}).

\bibitem{Sayre1952}
\bibinfo{author}{Sayre, D.}
\newblock \bibinfo{title}{Some implications of a theorem due to {Shannon}}.
\newblock \emph{\bibinfo{journal}{Acta Crystallogr.}}
  \textbf{\bibinfo{volume}{5},} \bibinfo{pages}{843--843}
  (\bibinfo{year}{1952}).

\bibitem{Fienup1982}
\bibinfo{author}{Fienup, J.}
\newblock \bibinfo{title}{Phase retrieval algorithms: a comparison}.
\newblock \emph{\bibinfo{journal}{Appl. Optics}} \textbf{\bibinfo{volume}{21},}
  \bibinfo{pages}{2758--2769} (\bibinfo{year}{1982}).

\bibitem{Miao1999}
\bibinfo{author}{Miao, J.}, \bibinfo{author}{Charalambous, P.},
  \bibinfo{author}{Kirz, J.} \& \bibinfo{author}{Sayre, D.}
\newblock \bibinfo{title}{Extending the methodology of {X-ray} crystallography
  to allow imaging of micrometre-sized non-crystalline specimens}.
\newblock \emph{\bibinfo{journal}{Nature}} \textbf{\bibinfo{volume}{400},}
  \bibinfo{pages}{342--344} (\bibinfo{year}{1999}).

\bibitem{Robinson1999}
\bibinfo{author}{Robinson, I.} \emph{et~al.}
\newblock \bibinfo{title}{Coherent {X-ray} diffraction imaging of silicon oxide
  growth}.
\newblock \emph{\bibinfo{journal}{Phys. Rev. B}} \textbf{\bibinfo{volume}{60},}
  \bibinfo{pages}{9965--9972} (\bibinfo{year}{1999}).

\bibitem{Seaberg2014}
\bibinfo{author}{Seaberg, M.} \emph{et~al.}
\newblock \bibinfo{title}{Tabletop nanometer extreme ultraviolet imaging in an
  extended reflection mode using coherent {Fresnel} ptychography}.
\newblock \emph{\bibinfo{journal}{Optica}} \textbf{\bibinfo{volume}{1},}
  \bibinfo{pages}{39} (\bibinfo{year}{2014}).

\bibitem{Zhang2015}
\bibinfo{author}{Zhang, B.} \emph{et~al.}
\newblock \bibinfo{title}{High contrast {3D} imaging of surfaces near the
  wavelength limit using tabletop {EUV} ptychography}.
\newblock \emph{\bibinfo{journal}{Ultramicroscopy}}
  \textbf{\bibinfo{volume}{158},} \bibinfo{pages}{98--104}
  (\bibinfo{year}{2015}).

\bibitem{Harada2013}
\bibinfo{author}{Harada, T.}, \bibinfo{author}{Nakasuji, M.},
  \bibinfo{author}{Nagata, Y.}, \bibinfo{author}{Watanabe, T.} \&
  \bibinfo{author}{Kinoshita, H.}
\newblock \bibinfo{title}{Phase imaging of extreme-ultraviolet mask using
  coherent extreme-ultraviolet scatterometry microscope}.
\newblock \emph{\bibinfo{journal}{Jpn. J. Appl. Phys.}}
  \textbf{\bibinfo{volume}{52},} \bibinfo{pages}{06GB02}
  (\bibinfo{year}{2013}).

\bibitem{Roy2011}
\bibinfo{author}{Roy, S.} \emph{et~al.}
\newblock \bibinfo{title}{Lensless {X-ray} imaging in reflection geometry}.
\newblock \emph{\bibinfo{journal}{Nature Photon.}}
  \textbf{\bibinfo{volume}{5},} \bibinfo{pages}{243--245}
  (\bibinfo{year}{2011}).

\bibitem{Sun2012}
\bibinfo{author}{Sun, T.}, \bibinfo{author}{Jiang, Z.},
  \bibinfo{author}{Strzalka, J.}, \bibinfo{author}{Ocola, L.} \&
  \bibinfo{author}{Wang, J.}
\newblock \bibinfo{title}{Three-dimensional coherent {X-ray} surface scattering
  imaging near total external reflection}.
\newblock \emph{\bibinfo{journal}{Nature Photon.}}
  \textbf{\bibinfo{volume}{6},} \bibinfo{pages}{588--592}
  (\bibinfo{year}{2012}).

\bibitem{Harder2013}
\bibinfo{author}{Harder, R.} \& \bibinfo{author}{Robinson, I.}
\newblock \bibinfo{title}{Coherent {X-ray} diffraction imaging of morphology
  and strain in nanomaterials}.
\newblock \emph{\bibinfo{journal}{JOM-US}} \textbf{\bibinfo{volume}{65},}
  \bibinfo{pages}{1202--1207} (\bibinfo{year}{2013}).

\bibitem{Robinson2009}
\bibinfo{author}{Robinson, I.} \& \bibinfo{author}{Harder, R.}
\newblock \bibinfo{title}{Coherent {X-ray} diffraction imaging of strain at the
  nanoscale}.
\newblock \emph{\bibinfo{journal}{Nature Mater.}} \textbf{\bibinfo{volume}{8},}
  \bibinfo{pages}{291--8} (\bibinfo{year}{2009}).

\bibitem{Jiang2010}
\bibinfo{author}{Jiang, H.} \emph{et~al.}
\newblock \bibinfo{title}{Quantitative {3D} imaging of whole, unstained cells
  by using {X-ray} diffraction microscopy}.
\newblock \emph{\bibinfo{journal}{P. Natl. Acad. Sci. USA}}
  \textbf{\bibinfo{volume}{107,}} \bibinfo{pages}{11234--9}
  (\bibinfo{year}{2010}).

\bibitem{Seibert2011}
\bibinfo{author}{Seibert, M.} \emph{et~al.}
\bibinfo{title}{Single mimivirus particles intercepted and imaged with an
  {X-ray} laser}.
\newblock \emph{\bibinfo{journal}{Nature}} \textbf{\bibinfo{volume}{470,}}
  \bibinfo{pages}{78--81} (\bibinfo{year}{2011}).

\bibitem{Dumas2013}
\bibinfo{author}{Dumas, C.}, \bibinfo{author}{van~der Lee, A.} \&
  \bibinfo{author}{Palatinus, L.}
\newblock \bibinfo{title}{Lensless coherent imaging of proteins and
  supramolecular assemblies: Efficient phase retrieval by the charge flipping
  algorithm}.
\newblock \emph{\bibinfo{journal}{J. Struct. Biol}}
  \textbf{\bibinfo{volume}{182,}} \bibinfo{pages}{106--16}
  (\bibinfo{year}{2013}).

\bibitem{Thibault2008}
\bibinfo{author}{Thibault, P.} \emph{et~al.}
\newblock \bibinfo{title}{High-resolution scanning {X-ray} diffraction
  microscopy.}
\newblock \emph{\bibinfo{journal}{Science}} \textbf{\bibinfo{volume}{321,}}
  \bibinfo{pages}{379--82} (\bibinfo{year}{2008}).

\bibitem{Maiden2009}
\bibinfo{author}{Maiden, A.} \& \bibinfo{author}{Rodenburg, J.}
\newblock \bibinfo{title}{An improved ptychographical phase retrieval algorithm
  for diffractive imaging}.
\newblock \emph{\bibinfo{journal}{Ultramicroscopy}}
  \textbf{\bibinfo{volume}{109,}} \bibinfo{pages}{1256--62}
  (\bibinfo{year}{2009}).

\bibitem{Corkum1993}
\bibinfo{author}{Corkum, P.}
\newblock \bibinfo{title}{Plasma perspective on strong field multiphoton
  ionization}.
\newblock \emph{\bibinfo{journal}{Phys. Rev. Lett.}}
  \textbf{\bibinfo{volume}{71,}} \bibinfo{pages}{1994--1997}
  (\bibinfo{year}{1993}).

\bibitem{Rundquist1998}
\bibinfo{author}{Rundquist, A.}
\newblock \bibinfo{title}{Phase-matched generation of coherent soft {X-rays}}.
\newblock \emph{\bibinfo{journal}{Science}} \textbf{\bibinfo{volume}{280,}}
  \bibinfo{pages}{1412--1415} (\bibinfo{year}{1998}).

\bibitem{Popmintchev2010}
\bibinfo{author}{Popmintchev, T.}, \bibinfo{author}{Chen, M.-C.},
  \bibinfo{author}{Arpin, P.}, \bibinfo{author}{Murnane, M.~M.} \&
  \bibinfo{author}{Kapteyn, H.~C.}
\newblock \bibinfo{title}{The attosecond nonlinear optics of bright coherent
  {X-ray} generation}.
\newblock \emph{\bibinfo{journal}{Nature Photon.}}
  \textbf{\bibinfo{volume}{4,}} \bibinfo{pages}{822--832}
  (\bibinfo{year}{2010}).

\bibitem{Hummler2013}
\bibinfo{author}{Hummler, K.} \emph{et~al.}
\newblock \bibinfo{title}{{TSV} and {Cu--Cu} direct bond wafer and
  package-level reliability}.
\newblock In \emph{\bibinfo{booktitle}{2013 IEEE 63rd Electronic Components and
  Technology Conference}}, \bibinfo{pages}{41--48} (\bibinfo{publisher}{IEEE},
  \bibinfo{address}{Las Vegas, Nevada, USA}, \bibinfo{year}{2013}).

\bibitem{Thibault2009}
\bibinfo{author}{Thibault, P.}, \bibinfo{author}{Dierolf, M.},
  \bibinfo{author}{Bunk, O.}, \bibinfo{author}{Menzel, A.} \&
  \bibinfo{author}{Pfeiffer, F.}
\newblock \bibinfo{title}{Probe retrieval in ptychographic coherent diffractive
  imaging}.
\newblock \emph{\bibinfo{journal}{Ultramicroscopy}}
  \textbf{\bibinfo{volume}{109,}} \bibinfo{pages}{338--43}
  (\bibinfo{year}{2009}).

\bibitem{Gardner2012}
\bibinfo{author}{Gardner, D.} \emph{et~al.}
\newblock \bibinfo{title}{High numerical aperture reflection mode coherent
  diffraction microscopy using off-axis apertured illumination}.
\newblock \emph{\bibinfo{journal}{Opt. Express}} \textbf{\bibinfo{volume}{20,}}
  \bibinfo{pages}{19050--9} (\bibinfo{year}{2012}).

\bibitem{Zhang2013}
\bibinfo{author}{Zhang, F.} \emph{et~al.}
\newblock \bibinfo{title}{Translation position determination in ptychographic
  coherent diffraction imaging}.
\newblock \emph{\bibinfo{journal}{Opt. Express}} \textbf{\bibinfo{volume}{21,}}
  \bibinfo{pages}{13592--606} (\bibinfo{year}{2013}).

\bibitem{Batey2014}
\bibinfo{author}{Batey, D.}, \bibinfo{author}{Claus, D.} \&
  \bibinfo{author}{Rodenburg, J.}
\newblock \bibinfo{title}{Information multiplexing in ptychography}.
\newblock \emph{\bibinfo{journal}{Ultramicroscopy}}
  \textbf{\bibinfo{volume}{138,}} \bibinfo{pages}{13--21}
  (\bibinfo{year}{2014}).

\bibitem{Windt1998}
\bibinfo{author}{Windt, D.~L.}
\newblock \bibinfo{title}{{IMD-Software} for modeling the optical properties of multilayer films}.
\newblock \emph{\bibinfo{journal}{Comput. Phys.}}
  \textbf{\bibinfo{volume}{12,}} \bibinfo{pages}{360} (\bibinfo{year}{1998}).

\bibitem{Moharam1995}
\bibinfo{author}{Moharam, M.}, \bibinfo{author}{Gaylord, T.},
  \bibinfo{author}{Pommet, D.} \& \bibinfo{author}{Grann, E.}
\newblock \bibinfo{title}{Stable implementation of the rigorous coupled-wave
  analysis for surface-relief gratings: enhanced transmittance matrix
  approach}.
\newblock \emph{\bibinfo{journal}{J. Opt. Soc. Am. A}}
  \textbf{\bibinfo{volume}{12,}} \bibinfo{pages}{1077} (\bibinfo{year}{1995}).

\bibitem{Nevot1980}
\bibinfo{author}{N{\'{e}}vot, L.} \& \bibinfo{author}{Croce, P.}
\newblock \bibinfo{title}{Caract{\'{e}}risation des surfaces par
  r{\'{e}}flexion rasante de rayons {X}. application {\`{a}} l'{\'{e}}tude du
  polissage de quelques verres silicates}.
\newblock \emph{\bibinfo{journal}{Rev. Phys. Appl.}}
  \textbf{\bibinfo{volume}{15,}} \bibinfo{pages}{761--779}
  (\bibinfo{year}{1980}).

\bibitem{Stearns1989}
\bibinfo{author}{Stearns, D.}
\newblock \bibinfo{title}{The scattering of {X-rays} from nonideal multilayer
  structures}.
\newblock \emph{\bibinfo{journal}{J. Appl. Phys.}}
  \textbf{\bibinfo{volume}{65,}} \bibinfo{pages}{491} (\bibinfo{year}{1989}).

\bibitem{Keil2007}
\bibinfo{author}{Keil, P.}, \bibinfo{author}{Lützenkirchen-Hech, D.} \& \bibinfo{author}{Frahm, R.}
\newblock \bibinfo{title}{Investigation of room temperature oxidation of {Cu}
  in air by {Yoneda}-{XAFS}}.
\newblock In \emph{\bibinfo{booktitle}{AIP Conference Proceedings}}, vol.
  \bibinfo{volume}{882,} \bibinfo{pages}{490--492} (\bibinfo{publisher}{AIP},
  \bibinfo{address}{Stanford, California, USA}, \bibinfo{year}{2007}).

\bibitem{Chase1998}
\bibinfo{author}{Chase, M.~J.}
\newblock \bibinfo{title}{{NIST-JANAF} thermochemical tables, 4th {Ed.}, {Part}
  {I}, {Al--Co}}.
\newblock \emph{\bibinfo{journal}{J. Phys. Chem. Ref. Data, Monograph No. 9}}
  (\bibinfo{year}{1998}).

\bibitem{Nakajima1997}
\bibinfo{author}{Nakajima, H.}
\newblock \bibinfo{title}{{The Discovery and Acceptance of the Kirkendall
  Effect: The Result of a Short Research Career}}.
\newblock \emph{\bibinfo{journal}{JOM-US}} \textbf{\bibinfo{volume}{49,}}
  \bibinfo{pages}{15--19} (\bibinfo{year}{1997}).

\bibitem{Hofmann2014}
\bibinfo{author}{Hofmann, S.}
\newblock \bibinfo{title}{Sputter depth profiling: past, present, and future}.
\newblock \emph{\bibinfo{journal}{Surf. Interface Anal.}}
  \textbf{\bibinfo{volume}{46,}} \bibinfo{pages}{654--662}
  (\bibinfo{year}{2014}).

\bibitem{Roberts1981}
\bibinfo{author}{Roberts, S.} \& \bibinfo{author}{Dobson, P.~J.}
\newblock \bibinfo{title}{Evidence for reaction at the {Al--SiO$_2$}
  interface}.
\newblock \emph{\bibinfo{journal}{J. Physics D Appl. Phys.}}
  \textbf{\bibinfo{volume}{14,}} \bibinfo{pages}{L17--L22}
  (\bibinfo{year}{1981}).

\bibitem{Zeng1983}
\bibinfo{author}{Zeng, L.}, \bibinfo{author}{Greibe, T.}, \bibinfo{author}{Nik,
  S.}, \bibinfo{author}{Delsing, P.} \& \bibinfo{author}{Olsson, E.}
\newblock \bibinfo{title}{Interdiffusion at the {Al/SiO$_2$} interface in
  {Al/AlOx/Al Josephson} junctions}.
\newblock In \emph{\bibinfo{booktitle}{Proc. of the 15th European Microscopy
  Congress}} (\bibinfo{address}{Manchester Central, United Kingdom},
  \bibinfo{year}{2012}).

\bibitem{Karl2015}
\bibinfo{author}{Karl, R.} \emph{et~al.}
\newblock \bibinfo{title}{Spatial, spectral, and polarization multiplexed
  ptychography}.
\newblock \emph{\bibinfo{journal}{Opt. Express}} \textbf{\bibinfo{volume}{23,}}
  \bibinfo{pages}{30250--8} (\bibinfo{year}{2015}).

\bibitem{Mortensen1995}
\bibinfo{author}{Mortensen, E.~N.} \& \bibinfo{author}{Barrett, W.~A.}
\newblock \bibinfo{title}{Intelligent scissors for image composition}.
\newblock In \emph{\bibinfo{booktitle}{Proc. of the 22nd annual conference on
  computer graphics and interactive techniques - SIGGRAPH `95}},
  \bibinfo{pages}{191--198} (\bibinfo{publisher}{ACM Press},
  \bibinfo{address}{New York, New York, USA}, \bibinfo{year}{1995}).

\end{thebibliography}
\end{document}